\documentclass[journal]{IEEEtran}
\usepackage{amsmath,amsfonts}
\usepackage{algorithmic}
\usepackage{algorithm}
\usepackage{array}
\usepackage[caption=false,font=footnotesize,labelfon
t=sf,textfont=sf]{subfig}
\usepackage{cite}
\usepackage{textcomp}
\usepackage{stfloats}
\usepackage{url}
\usepackage{verbatim}
\usepackage{graphicx}
\usepackage{cite}
\usepackage{bm}
\usepackage{amssymb}

\usepackage{amsfonts}
\usepackage{amsmath}
\usepackage{empheq}

\usepackage{graphicx}
\usepackage{multirow}
\usepackage{gensymb}
\usepackage{graphicx}
\usepackage{epsfig}
\usepackage{svg}
\usepackage{ragged2e} 
\usepackage[colorlinks,linkcolor=blue,citecolor=blue]{hyperref}

\usepackage{caption}

\begin{document}

\title{Beyond Traditional Coherence Time: An  Electromagnetic Perspective for Mobile Channels}

\author{Zihan Zhou, Li Chen,~\IEEEmembership{Senior Member}, Ang Chen, and Weidong Wang
\thanks{The authors are with the CAS Key Laboratory of Wireless-Optical
Communications, University of Science and Technology of China, Hefei
230027, China (e-mail: zzh31@mail.ustc.edu.cn; chenli87@ustc.edu.cn; chenang1122@mail.ustc.edu.cn; wdwang@ustc.edu.cn)}
 \vspace{-0.42cm}
} 
\markboth{Journal of \LaTeX\ Class Files,~Vol.~m, No.~n, March~2025}%
{Shell \MakeLowercase{\textit{et al.}}: A Sample Article Using IEEEtran.cls for IEEE Journals}


\maketitle
\begin{abstract} 
Channel coherence time has been widely regarded as a critical parameter in the design of mobile systems. However, a prominent challenge lies in integrating electromagnetic (EM) polarization effects into the derivation of the channel coherence time.
In this paper, we develop a framework to analyze the impact of polarization mismatch on the channel coherence time.
Specifically, we first establish an EM channel model to capture the essence of EM wave propagation. Based on this model, we then derive the EM temporal correlation function, incorporating the effects of polarization mismatch and beam misalignment. Further, considering the random orientation of the mobile user equipment (UE), we derive a closed-form solution for the EM coherence time in the turning scenario. When the trajectory degenerates into a straight line, we also provide a closed-form lower bound on the EM coherence time. The simulation results validate our theoretical analysis and reveal that neglecting the EM polarization effects leads to overly optimistic estimates of the EM coherence time.

\end{abstract}

\vspace{-0.1cm}
\section{Introduction}
\IEEEPARstart{N}{ext} generation wireless networks are expected to support emerging applications with complex mobility patterns, including autonomous driving (AD), virtual reality (VR), and augmented reality (AR) \cite{b}. A prominent challenge here is the link instability caused by rapid channel fading due to high mobility. To capture the dynamic channel variations, the channel coherence time is regarded as a crucial factor determining the update frequency of channel state information (CSI) \cite{c}. Within a coherence time, the channel response is considered constant \cite{d}. Thus, it is of practical importance to properly quantify the channel coherence time.

Numerous studies have proposed various methods to determine the channel coherence time. A rule of thumb sets it as the duration required for the user equipment (UE) to move a quarter-wavelength distance \cite{e}, while a more rigorous definition specifies it as the maximum time interval maintaining channel temporal correlation above a predefined threshold \cite{f}. Furthermore, the authors of \cite{g} related the coherence time with the beam pointing error due to UE mobility and proposed a novel metric called the beam coherence time, which is an order of magnitude longer than the conventional coherence time. The work in \cite{h} investigated the impact of UE mobility under non-rotational trajectories and derived a closed-form expression for the worst-case coherence time.

Note that the above works ignored certain electromagnetic (EM) effects in channel modeling, specifically the polarization mismatch.
In \cite{j}, the authors indicated that the impact of polarization mismatch could significantly influence the temporal dynamics of the mobile channel. A more accurate approach to evaluating the coherence time should be proposed based on the channel model using the EM theory. Recently, the EM channel, derived from Maxwell's equations, has been extensively investigated. The works in \cite{k} and \cite{l} utilized the Green’s function to model the EM channel, which captures the EM propagation characteristics. To further leverage the above theoretical progress in practical system design, the authors of \cite{d1} proposed a new channel estimation scheme based on the EM theory.
However, there is no prior work that integrates EM polarization effects into the coherence time determination. Unlike conventional channels used in traditional coherence time analyses, EM channels involve more complex amplitude and phase modeling, which is mathematically intractable \cite{m}. This makes it challenging to derive the closed-form expression for the EM coherence time.


Motivated by the above discussion, in this paper, we develop an EM channel model to accurately characterize the impact of polarization mismatch. Then, accounting for both the beam misalignment and variations in the EM polarization effects caused by UE mobility, we present the temporal correlation function of the EM channel. Further, we consider two typical motion scenarios, i.e., the turning and linear motion scenarios. For these two scenarios, we derive the closed-form expressions for the EM coherence time.  Simulation results validate our analytical solutions and demonstrate that neglecting polarization mismatch leads to over-optimistic coherence time estimates.

\section{System Model}
\begin{figure}
\centering
\includegraphics[width=3in]{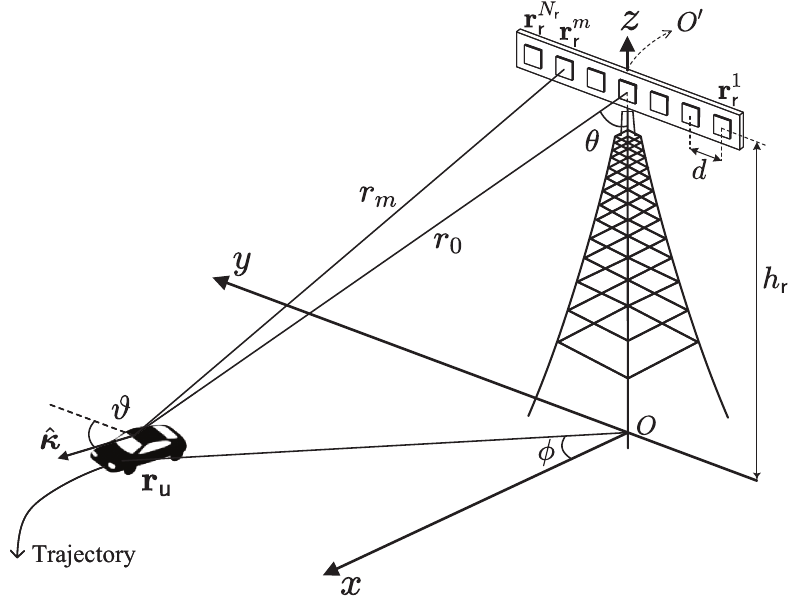}
\caption{Illustration of the cellular mobile system model at time $t$, where the time index $t$ is ignored for notational convenience.}
\label{fig_1}
\vspace{-0.2cm} 
\end{figure}

As depicted in Fig. \ref{fig_1}, we consider a cellular mobile system, where
a base station (BS) equipped with a uniform linear array (ULA) consisting of $N_\mathsf{r}$ patch antenna elements
serves a mobile UE configured with a uni-polarized antenna.

We establish an orthonormal coordinate system $O$-$xyz$ with the center of the ULA positioned at point $O^\prime$ along the $z$-axis.
The ULA elements are configured along the $x$-axis with inter-element spacing $d$, and $h_\mathsf{r}$ is the height of the array. The position of the $m$-th receiving antenna can be denoted as $\mathbf{r}^m_{\mathsf{r}} = \left[0, y^m_{\mathsf{r}}, h_\mathsf{r}\right]^\mathsf{T}$, where $ y^m_{\mathsf{r}}=\delta_m d$ and $\delta_m = \frac{1}{2}(2m-N_{{\mathsf{r}}}+1)$. 
The position of the UE at time $t$ is denoted as $\mathbf{r}_{\mathsf{u}}(t)=\left[x_{\mathsf{u}}(t), y_{\mathsf{u}}(t), 0\right]^\mathsf{T}$. 
Then, the distance between the UE and the $m$-th antenna  can be calculated as 
\begin{align}
\notag r_m(t)&=\left\|\mathbf{r}_m(t)\right\|=\left\|\mathbf{r}^m_{\mathsf{r}}-\mathbf{r}_{\mathsf{u}}(t)\right\| \\& =\sqrt{r^2(t)+\delta_m^2 d^2-2 r(t)\delta_m d \sin \phi(t) \sin \theta(t)}
\ ,
\label{A}
\end{align}
where $r(t)$ is the BS-UE distance between the center of the ULA and the UE, $\phi(t)$ and $\theta(t)$ denote the azimuth and elevation components of the angles of arrival (AoAs),  respectively.


To describe the motion state, we consider that the UE moves at a constant speed $v$  with its antenna orientation changing in accordance with the direction of motion \cite{z}. We define a normalized vector $\hat{\boldsymbol{\kappa}}(t)=\left[\sin\vartheta(t), \cos\vartheta(t), 0\right]^\mathsf{T}$ to indicate the instantaneous orientation of the transmit antenna, where $\vartheta(t)$ denotes the heading angle with respect to the $y$-axis. Within a short duration, the turn rate $\omega$ can be considered constant, and the trajectory can be modeled as a circular arc segment with the turning radius  $\rho=v/\omega$. 
\vspace{-0.3cm}
\subsection{EM Channel Model}
To accurately capture the EM polarization effects of wireless signals, we derive the channel model from the EM perspective, where the source current of the UE generates the radiated electric field at the BS to establish the communication link. Specifically, the phasor of the current density along the UE can be written as $\boldsymbol{\mathcal{J}}\left(t\right)=I_0\hat{\boldsymbol{\kappa}}(t)$, where $I_0$ is the scalar current density.
Based on Maxwell’s equations, the resulting electric field at the position $\mathbf{r}^m_{\mathsf{r}}$ is
\begin{align}
    \boldsymbol{\mathcal{E} }\left(t\right)=\int_{S_{\mathsf{u}}} \boldsymbol{\mathcal{G}}_m\left(t\right) \boldsymbol{\mathcal{J}}\left(\mathbf{r}_{\mathsf{u}}(t)\right)\mathrm{d} \mathbf{r}_{\mathsf{u}}(t),
    \label{B}
        \vspace{-0.1cm} 
\end{align}
where $S_{\mathsf{u}}$ is the source region and $\boldsymbol{\mathcal{G}}_m\left(t\right)\in \mathbb{C}^{3 \times 3}$ is the dyadic Green’s function. Under the assumption $r_m(t)\gg\lambda$, where $\lambda$ is the wavelength, $\boldsymbol{\mathcal{G}}_m\left(t\right)$ can be approximated as \cite{n}
\begin{align}
\boldsymbol{\mathcal{G}}_m\left(t\right)
& \simeq \frac{j \eta}{2 \lambda r_m(t)} e^{j k r_m(t)}\left(\mathbf{I}-\hat{\mathbf{r}}_m(t) \hat{\mathbf{r}}^\mathsf{T}_m(t)\right),
\label{C}
\end{align}
where $j=\sqrt{-1}$, $\mathbf{I}$ indicates the identity matrix, $\hat{\mathbf{r}}_m(t)=\mathbf{r}_m(t) /\|\mathbf{r}_m(t)\|$ is the unit vector of ${\mathbf{r}_m}$, $\eta$ is the intrinsic impedance of free space, and $k = 2\pi/\lambda$ is the wavenumber. Substituting \eqref{C} into \eqref{B}, we have 
\begin{align}
\boldsymbol{\mathcal{E} }\left(t\right) & =  j\frac{\mathcal{V}_{\mathsf{in }}}{\lambda r_m(t)} e^{j k r_m(t)}\left(\hat{\mathbf{r}}_m(t) \otimes \hat{\boldsymbol{\kappa}}(t)\right) \otimes \hat{\mathbf{r}}_m(t),
\label{D}
\end{align}
where $\otimes$ is the cross product and $\mathcal{V}_{\mathsf{in }}=\frac{\eta}{2}I_0S_{\mathsf{u}}$ is the initial voltage. The EM channel between the UE and the $m$-th antenna can be modeled as 
\begin{align}
h_{m}(t)=\beta_m(t) e^{j k r_m(t)},
\end{align}
where the channel amplitude $\beta_m(t)$ is given by
\begin{align}
\notag
&\beta_m(t)=  \frac{\| \boldsymbol{\mathcal{E}}\left(t\right) \|}{\mathcal{V}_\mathsf{in}} \sqrt{A_{\mathsf{t}}}\sqrt{-\hat{\mathbf{r}}_m^\mathsf{T}(t) \hat{\mathbf{u}}_x} \\
& = \sqrt{\underbrace{\frac{1}{4 \pi r_m^2(t)}}_{\text {FSPL}} \underbrace{\frac{h_{\mathsf{r}}^2+r_{\mathsf{h}, m}^2(t) \cos ^2(\vartheta(t)+\phi(t))}{r_m^2(t)}}_{\text {Polarization mismatch loss}} \underbrace{\frac{x_{\mathsf{u}}(t)}{r_m(t)}}_{\text {EA loss }}},
 \label{F}
\end{align}
where $r_{\mathsf{h},m}(t)=\sqrt{x_{\mathsf{u}}^2(t)+\left[y_{\mathsf{r}}^m-y_{\mathsf{u}}(t)\right]^2}$ is the projection of $r_{m}(t)$ onto the horizontal plane, $\hat{\mathbf{u}}_x$ is a unit vector along the $x$-axis, $\sqrt{-\hat{\mathbf{r}}_m^\mathsf{T} \hat{\mathbf{u}}_y}$ is the projection coefficient \cite{o}, and $A_{\mathsf{t}}=\lambda^2 /(4 \pi)$ is the effective area of the isotropic antenna\cite{p}. 
Unlike the conventional free-space path loss (FSPL) model, the channel amplitude in \eqref{F} incorporates the polarization mismatch loss and effective aperture (EA) loss.

Since the distance $r_m(t)$ is much larger than the inter-element spacing, the amplitude variations across array elements can be neglected, i.e., $\beta_{m}(t) \simeq \beta_{0}(t), \forall m$ with $\beta_{0}(t)$ denoting the channel amplitude at the center of the array. For notational brevity, we omit the subscript $0$ for further discussion. Then, $h_{m}(t)$ can be extended to a multi-antenna channel vector as 
\begin{align}
\mathbf{h}(t)  = \beta(t) \mathbf{a}_{\mathsf{r}}\left[\phi(t), \theta(t) \right]e^{jkr(t)},
 \label{FF}
\end{align}
where $\mathbf{a}_{\mathsf{r}}\left[\phi(t), \theta(t) \right]\in \mathbb{C}^{N_\mathsf{r} \times 1}$ is the steering vector. Using the Fresnel approximation\cite{y}, the $n$-th element of $\mathbf{a}_{\mathsf{r}}\left[\phi(t), \theta(t) \right]$ can be written as 
\begin{align}
\left\{\mathbf{a}_{\mathsf{r}}\left[\phi(t), \theta(t) \right]\right\}_n=e^{-j k d \left[\delta_n \Psi(t)-\frac{\delta_n^2 d^2\left(1-\Psi^2(t)\right)}{2 r(t)}\right]}, 
\end{align}
where $\Psi(t)=\sin \phi(t) \sin \theta(t)$. 

\subsection{EM Temporal Correlation Function } 

The mobility of the UE leads to the channel's time-varying characteristics.
To mitigate the effects of channel mismatch due to the motion, the UE is required to transmit pilot signals after each EM coherence time $T_{\mathsf{EM}}$ to update the CSI. During each $T_{\mathsf{EM}}$, the channel response remains approximately constant. Nevertheless, 
 a non-zero delay $\tau$ inevitably exists between the estimated channel and the actual channel within the $T_{\mathsf{EM}}$.
 
To derive the EM coherence time, we consider the EM temporal correlation function, which is denoted as
\begin{align}
&R(\tau)  \notag =\mathbb{E}\left[\frac{\left|\mathbf{h}^\mathsf{H}(0) \mathbf{h}(\tau)\right|}{\operatorname{max}\left\{\|\mathbf{h}(0)\|^2,\|\mathbf{h}(\tau)\|^2\right\}} \right] \\
&\!=\!\mathbb{E}\left[\frac{\operatorname{min}\{\beta(0), \beta(\tau)\}}{\operatorname{max}\{\beta(0), \beta(\tau)\}} \frac{\left|\mathbf{a}_{\mathsf{r}}^\mathsf{H}[\phi(0), \theta(0)] \mathbf{a}_{\mathsf{r}}[\phi(\tau), \theta(\tau)]\right|}{N_\mathsf{r}}\right]\!\!,
 \label{I}
\end{align}
where $R(\tau)$ is comprised of two components. The first component characterizes the channel amplitude variations, which capture the polarization mismatch. The second component represents the beam misalignment resulting from AoA errors induced by UE mobility. 

Based on the EM temporal correlation function in \eqref{I}, we define the EM coherence time $T_{\mathsf{EM}}$ as 
\begin{align}
T_{\mathsf{EM}}=\inf _\tau\left\{\tau \left\lvert\, R(\tau)<\zeta\right.\right\},
 \label{J}
\end{align}
where  $\zeta$ is the specified threshold. Using this definition, we derive the EM coherence time in the next section.

\section{Electromagnetic Coherence Time}
In this section, we first analyze the impact of UE motion on AoAs variations using a geometric model. Taking both the above angle variations and the polarization mismatch into consideration, we then derive the closed-form expressions for the EM coherence time for two specific scenarios, i.e., the turning scenario and linear motion scenario.
 
\subsection{AoAs Variations due to UE Mobility}

\begin{figure}
\centering
\includegraphics[width=3.2in]{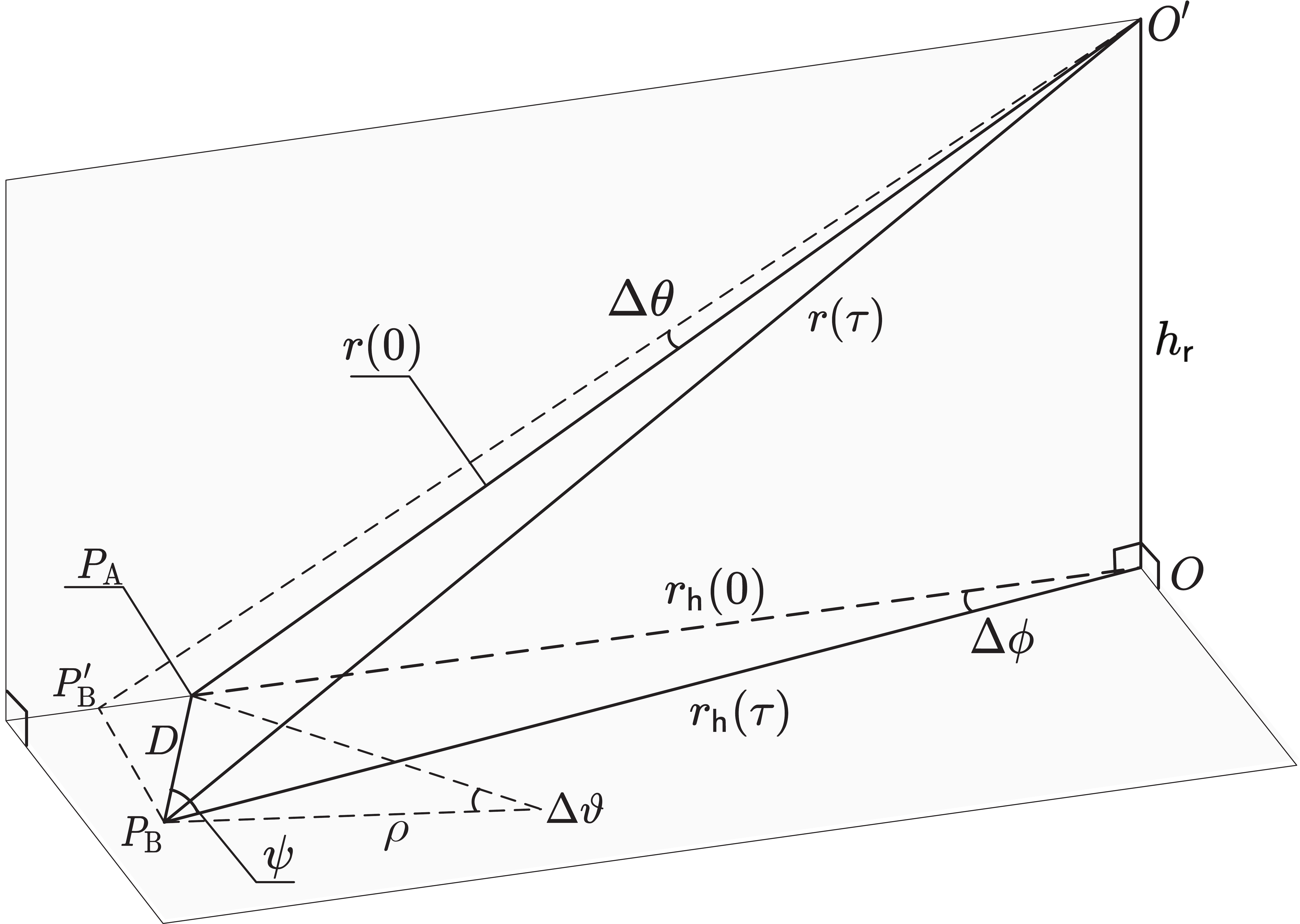}
\caption{The geometric model when UE moves within a duration $\tau$.}
\label{fig_2}
\vspace{-0.4cm} 
\end{figure}

As shown in Fig. \ref{fig_2}, we assume that the UE is located at point $P_\mathrm{A}$ at time $0$ and arrives at point $P_\mathrm{B}$ at time $\tau$. The variation in $\vartheta(t)$ within $\tau$ is represented as $\Delta\vartheta=\omega \tau$ and the displacement between $P_\mathrm{A}$ and $P_\mathrm{B}$ is $D\!=\!2 \rho \sin \frac{\Delta \vartheta}{2}$. We denote the projection of $r(t)$ onto the horizontal plane as $r_{\mathsf{h}}(t)$, point $P_\mathrm{B}^{\prime}$ lies on the extension of line $OP_\mathrm{A}$ and satisfies $OP_\mathrm{B}\!=\!OP_\mathrm{B}^\prime$. Additionally, we define the angle $\psi=\angle O P_\mathrm{B} P_\mathrm{A}$. 

According to the geometric relationship, the variation in azimuth angle $\Delta \phi$ within a duration $\tau$ can be derived as
\begin{align}
\Delta \phi \simeq \frac{D\sin \psi}{r_{\mathsf{h}}(0)}  \simeq \frac{2 \rho}{r_{\mathsf{h}}(0)} \sin \psi \sin \frac{\Delta \vartheta}{2},
\label{L}
\end{align}
where we use the approximation $\sin \Delta \phi \simeq \Delta \phi$ for small $\Delta \phi$. 
Since $\sin \Delta \theta \simeq \Delta \theta$ and $r_{\mathsf{h}}(\tau)-r_{\mathsf{h}}(0) \simeq D \cos \psi$, the variation in elevation angle $\Delta\theta$ can be derived as
\begin{align}\notag
\Delta \theta &\simeq \frac{r_{\mathsf{h}}(\tau)-r_{\mathsf{h}}(0)}{r(0)} \sin \angle O P^\prime_\mathrm{B} O^{\prime}\\
&\simeq  \frac{2\rho}{r(0)} \cos \psi \frac{h_{\mathsf{r}}}{r(\tau)} \sin \frac{\Delta \vartheta}{2}.
\label{TT}
\end{align}

It can be observed that the angular variations within $\tau$ are bounded by $\Delta \phi\leq\frac{2\rho}{r_{\mathsf{h}}(0)}$ and  $\Delta \theta\leq\frac{2\rho h_\mathsf{r}}{r(0)r(\tau)}$. 
Considering the above motion-induced angle errors $\Delta \phi$ and $\Delta \theta$, $R(\tau)$ in \eqref{I} can be further expressed as a complicated function, which presents challenges in calculating the EM coherence time $T_{\mathsf{EM}}$. To derive a closed-form expression of $T_{\mathsf{EM}}$, we consider two typical motion scenarios: the turning scenario and the linear motion scenario.

\vspace{-0.3 cm}
\subsection{EM Coherence Time for Turning Scenario}
In this scenario, the UE travels a curve with a small turning radius $\rho \ll r_{\mathsf{h}}(0)$, leading to $\Delta \phi\rightarrow0$ and $ \Delta \theta \rightarrow 0$. 
This implies that the impact of beam misalignment is negligible. Since $\rho$ is small, the displacement has a negligible effect on the channel amplitude, and the UE's position can be considered unchanged within $\tau$, i.e., $x_\mathsf{u}(\tau)\simeq x_\mathsf{u}(0)$ and $y_\mathsf{u}(\tau)\simeq y_\mathsf{u}(0)$.

When the UE is turning, the orientation of the antenna changes accordingly \cite{z}, and the variation of the polarization mismatch loss exerts a pronounced impact on channel temporal correlation. Assuming that the initial heading angle $\vartheta_0\!\triangleq\!\vartheta(0)$ is uniformly distributed in $[0, 2\pi]$, $R(\tau)$ in \eqref{I} is simplified to 
\begin{align}
 R(\tau) =\frac{1}{2 \pi} \int_{\mathcal{C}}\left[\frac{\beta(\tau)}{\beta(0)} \mathbb{I}_{\mathcal{A}}+\frac{\beta(0)}{\beta(\tau)} \mathbb{I}_{\mathcal{B}}\right]\mathrm{d} \vartheta_0 \!=\!\frac{R_{\mathcal{A}}+R_{\mathcal{B}}}{2 \pi},
\label{N}
\end{align}
where $\mathbb{I}$ is the indicator function, $R_\mathcal{A} =\int_{\mathcal{A}}\frac{\beta(\tau)}{\beta(0)}  \mathrm{d} \vartheta_0$ and $R_\mathcal{B}  =\int_{\mathcal{B}}\frac{\beta(0)}{\beta(\tau)}  \mathrm{d} \vartheta_0$ are two integrals of $R(\tau)$. Due to the periodicity of trigonometric functions, $\mathcal{C}$ can be taken as any continuous interval of length $2 \pi$. $\mathcal{A}$ is defined as the region within $\mathcal{C}$ where $\frac{\beta(\tau)}{\beta(0)}<1$ and $\mathcal{B}$ is the complement of $\mathcal{A}$ within $\mathcal{C}$.

\vspace{0.1cm}
By substituting the channel amplitude in \eqref{F} into \eqref{N} and applying the small $\rho$ approximation, $R_\mathcal{A}$ can be written as
\begin{align}
\notag R_\mathcal{A} &\simeq\int_{\mathcal{A}}\sqrt{\frac{h_{\mathsf{r}}^2+r_{\mathsf{h}}^2(0) \cos ^2(\vartheta_0+\Delta\theta+ \phi(0))}{h_{\mathsf{r}}^2+r_{\mathsf{h}}^2(0) \cos ^2\left(\vartheta_0+\phi(0)\right)}}\mathrm{d} \vartheta_0
\\
& = \int_{\mathcal{A}} \sqrt{1+\frac{\cos \left( \vartheta^{\prime}_0+2 \Delta \theta\right)-\cos\vartheta^{\prime}_0}{1+\frac{2 h^2}{{r}_{\mathsf{h}}^2(0)}+\cos \vartheta^{\prime}_0}} \mathrm{d} \vartheta_0,
\label{O}
\end{align}
where $\vartheta^{\prime}_0=2 \vartheta_0+2 \phi(0)$. 
Based on the above definition, $\mathcal{A}$ is determined by the inequality condition $\cos \left( \vartheta^{\prime}_0+2 \Delta \theta\right)-\cos\vartheta^{\prime}_0<0$. Thus, $\mathcal{A}$ can be calculated as 
\begin{align}
\mathcal{A} =\left[\alpha_0, \alpha_1\right] \cup\left[\alpha_2, \alpha_3\right] \triangleq\bigcup_{i=1}^2\left[a_1^i, a_2^i\right],
\label{Na}
\end{align}
where  $\alpha_k=\frac{1}{2} k \pi-\phi(0)-\frac{1}{2} \Delta \vartheta$, $a_1^i$ and $a_2^i$ denote the lower and upper bounds of the $i$-th sub-interval, respectively.

Using $\sqrt{1+z}\simeq1+\frac{z}{2}$ for small $z$ and $\cos (\theta+\Delta \theta) \simeq \cos \theta- \Delta \theta \sin \theta$ for small $\Delta \theta$, \eqref{O} can be rewritten as
\begin{align}
\notag R_\mathcal{A} &=\int_\mathcal{A} 1-\frac{\Delta \vartheta \sin \vartheta_0^{\prime}}{1+\frac{2 h_\mathsf{r}^2}{{r}_{\mathsf{h}}^2(0)}+\cos \vartheta_0^{\prime}} \mathrm{d} \vartheta_0\\
\notag&=\left.\sum_{i=1}^2\left[ \vartheta_0\!+\!\frac{\Delta \vartheta }{2}\ln\! \left(1\!+\!\frac{2 h_\mathsf{r}^2}{{r}_{\mathsf{h}}^2(0)}\!+\!\cos (2 \vartheta_0\!+ \!2\phi(0))\right)\right]\right|_{a_1^i}^{a_2^i}.
\end{align}

Similarly, the region $\mathcal{B}$ can be represented as $\left[\alpha_1, \alpha_2\right] \cup\left[\alpha_3, \alpha_4\right]\triangleq\bigcup_{i=1}^2\left[b_1^i, b_2^i\right]$, and the second integral $R_\mathcal{B}$ is also obtained using the same approach as 
\begin{align}
\notag R_\mathcal{B} \left.\!=\!\!\sum_{i=1}^2\!\left[\vartheta_0\!-\!\frac{\Delta \vartheta}{2}\! \ln \!\left(\!1\!+\!\frac{2 h_{\mathsf{r}}^2}{{r}_{\mathsf{h}}^2\!(0)}\!+\!\cos (2\vartheta_0\!+\!2 \phi(0)\!+\!2 \Delta \vartheta)\!\right)\!\right]\right|_{b_1^i}^{b_2^i}\!\!.
\end{align}

Therefore, the correlation function for the turning scenario can be derived as 
\begin{align}
\notag R(\tau) &=1-\frac{\Delta \vartheta}{ \pi} \ln\left(\frac{1+\frac{2 h_\mathsf{r}^2}{{r}_{\mathsf{h}}^2(0)}+\cos \Delta \vartheta}{1+\frac{2 h_\mathsf{r}^2}{{r}_{\mathsf{h}}^2(0)}-\cos \Delta \vartheta}\right) \\
&\overset{(a)}{\simeq} 1-\frac{\Delta \vartheta}{\pi} \ln \frac{4}{(\Delta \vartheta)^2},
\end{align}
where the step $(a)$ holds when ${r}_{\mathsf{h}}^2(0)$ is large and $\frac{1+\cos \Delta \vartheta}{1-\cos \Delta \vartheta} \simeq \frac{4}{(\Delta \vartheta)^2}$ for small $\Delta \vartheta$. Following the definition in \eqref{J}, we can \noindent derive the closed-form solution of the EM coherence time in the turning scenario as
\begin{align}
T_{\mathsf{EM}}=-\frac{\pi(1-\zeta) \rho}{2 v \mathcal{W}_{-1}\left(-\frac{(1-\zeta) \pi}{4}\right)},
\label{S}
\end{align}
where $\mathcal{W}_{-1}\left(\cdot\right)$ is a branch of the Lambert W function\cite{w}. Note that for a given threshold $\zeta$, the EM coherence time in the turning scenario is directly proportional to the turn rate $v/\rho$ and is independent of the distance.

\subsection{EM Coherence Time for Linear Motion Scenario}

When the turning radius goes to infinity, the motion model degenerates into a linear motion scenario. Since the direction of travel remains constant, the impact of polarization mismatch can be negligible. Then, by ignoring EM polarization effects, the correlation function in \eqref{I} can be simplified as:
\begin{align}
&\notag R(\tau) =\mathbb{E}\left[\frac{\left| \mathbf{a}_{\mathsf{r}}^{\mathsf{H}}[\phi(0), \theta(0)] \mathbf{a}_{\mathsf{r}}[\phi(\tau), \theta(\tau)]\right|}{N_{\mathsf{r}}}\right] \\
& \!=\! \mathbb{E}\left[\frac{1}{N_{\mathsf{r}}} \left|\sum_{n=0}^{N_{\mathsf{r}}-1}e^{-j k d \delta_n\left(\Psi(\tau)-\Psi(0)-d \delta_n \frac{\Psi(\tau)^2-\Psi^2(0)}{2 r(0)}\right)}\right|\right].
\label{T}
\end{align}
Note that it is hard to obtain a closed-form expression for $T_\mathsf{EM}$ from \eqref{T}. To obtain precise results, we solve it numerically. 

Further, we consider a more typical case, where $r(0)$ is large. In this case, the quadratic phase term of $\mathbf{h}(t)$ in \eqref{FF} is neglected. $\Delta \theta$ in \eqref{TT}, as a higher
order term with respect to $\frac{1}{r(0)}$, can also be negligible. Therefore, the correlation function with specific AoAs can be expressed as 
\begin{align}
\notag R(\tau)&=\left|\frac{1}{N_{\mathsf{r}}} \sum_{n=0}^{N_{\mathsf{r}}-1} e^{-j k d \delta_n \sin \theta(0)[\sin (\phi(0)+\Delta \phi)-\sin \phi(0)]}\right|\\ \notag 
& \overset{(b)}{\simeq} e^{-\frac{N_{\mathsf{r}}^2}{2} \sin \theta(0)[\sin (\phi(0)+\Delta \phi)-\sin \phi(0)]^2}\\
& \overset{(c)}{\simeq} e^{-\frac{N_{\mathsf{r}}^2}{2} \sin \theta(0)\left(\frac{\sin \psi \cos\phi(0)}{r_{\mathsf{h}}(0)} v \tau\right)^2},
\label{U}
\end{align}
where $(b)$ uses  
$\frac{1}{N_{\mathsf{r}}^2}\left|\sum_{m=0}^{N_{\mathsf{r}}-1} e^{-j \pi m \Delta}\right| \simeq e^{-\frac{N_{\mathsf{r}}^2}{2} \Delta^2}$ for small $\Delta$\cite{q}, $(c)$ uses $\sin(\phi+\Delta \phi)-\sin \phi \simeq \Delta \phi \cos \phi$ for small $\Delta \phi$, the displacement $D=\mathnormal{v} \tau$ and we assume $d=\lambda/2$.
Following the definition in \eqref{J}, we obtain a closed-form expression for the EM coherence time for the linear motion scenario as 
\begin{align}
T_{\mathsf{EM}}=\frac{r_{\mathsf{h}}(0)}{v \left|\sin \psi \cos\phi(0)\right|} \sqrt{\frac{2\ln (1/\zeta)}{N_\mathsf{r}^2 \sin \theta(0)}}.
\label{V}
\end{align}
Since a robust EM coherence time should account for the worst-case scenario, the lower bound is practically significant in protocol design. Given that $\left|\sin \psi \cos\phi(0)\right|<1$, the EM coherence time in \eqref{V} can be lower bounded by
\begin{align}
T_{\mathsf{EM},\mathsf{LB}}=\frac{r_{\mathsf{h}}(0)}{v} \sqrt{\frac{2\ln (1/\zeta)}{N_\mathsf{r}^2 \sin \theta(0)}},
\label{W}
\end{align}
which is related explicitly to the distance $r_{\mathsf{h}}(0)$, the speed $v$, and the number of antennas $N_\mathsf{r}$. 

\section{Simulation Results}

In this section, we present numerical results to show the impact of UE mobility on the EM coherence time and verify our derivations for turning and linear motion scenarios. The system is operated at $28$ GHz with a wavelength of $\lambda = 0.125$ m. The BS is equipped with $N_{\mathsf{r}} = 64$ antennas and its height is $h_{\mathsf{r}} = 3$ m. The UE moves within an activity range of $r\in$$[10$ m, $300$ m$]$ and the threshold is set to $\zeta\!=\!0.9$. In each experiment, we assume perfect knowledge of the azimuth angle at $t= 0$, e.g., $\phi(0) = 90$$\degree$. The initial heading angle follows a uniform distribution, i.e., $\vartheta_0 \sim \! \mathcal{U}\left[0, {2\pi}\right]$. The solutions are obtained through $20,000$ times Monte-Carlo trials.

\begin{figure}[h]
\centering\hspace{-1 mm}
\includegraphics[width=3in]{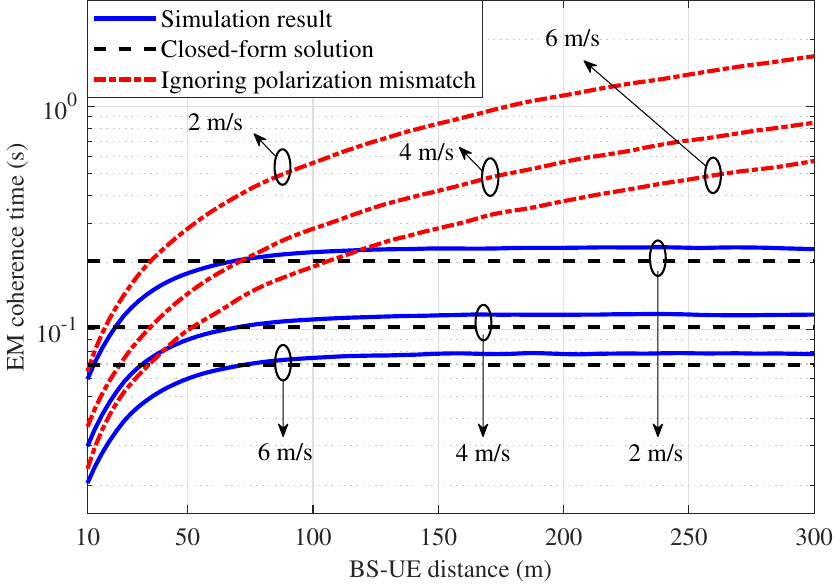}
\caption{EM coherence time v.s. BS-UE distance in the turning scenario.}
\label{fig_3}
\end{figure}

Fig. \ref{fig_3} illustrates the EM coherence time v.s. the BS-UE distance when the UE moves with different speeds $v = \left\{2,4,6\right\}$ m/s and a turning radius $\rho = 10$ m. The simulation result is obtained based on the correlation function in \eqref{I}, the closed-form solution is derived in \eqref{S}, and the result obtained by ignoring polarization mismatch is the EM coherence time calculated numerically from the simplified correlation function in \eqref{T}. It can be observed that the simulation result initially increases with distance and reaches a stationary value, which can be approximated by the closed-form solution. This validates the solution in \eqref{S} as an effective computational scheme for EM coherence time. Notably, neglecting the impact of polarization mismatch leads to a significant overestimation of the EM coherence time by an order of magnitude due to the pronounced effect of polarization mismatch in the turning scenario. 
\begin{figure}[!t]
\centering\hspace{-3 mm}
\includegraphics[width=3in]{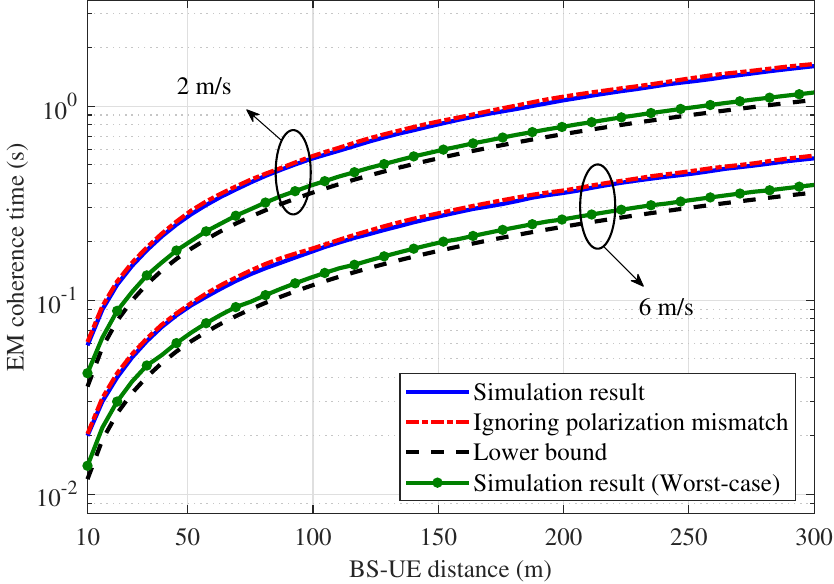}
\caption{EM coherence time v.s. BS-UE distance in the linear motion scenario.}
\label{fig_4}
\vspace{-0.1cm}
\end{figure}

Fig. \ref{fig_4} depicts the EM coherence time v.s. the BS-UE distance when the UE moves in a fixed direction. To validate the lower bound in \eqref{W}, we consider the worst-case scenario $\left( \vartheta_0 = 90 \degree\right)$, in which the correlation function experiences the fastest fading, yielding the shortest EM coherence time. It can be observed that over the range considered, the lower bound in \eqref{W} remains valid. Besides, it can be seen that neglecting the impact of polarization mismatch has a negligible impact on the EM coherence time calculation. This is because the orientation of the transmit antenna remains unchanged despite the motion of the UE. 

\begin{figure}[!t]
\centering\hspace{-1 mm}
\includegraphics[width=3in]{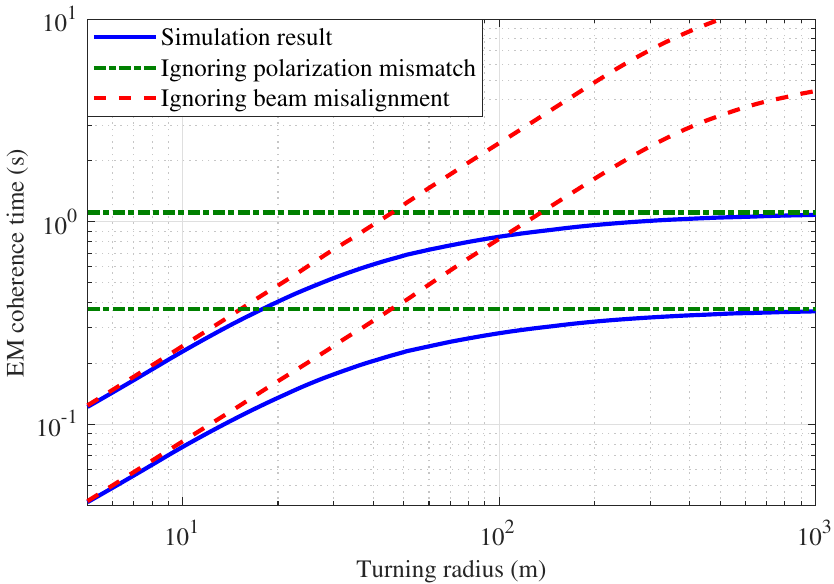}
\caption{The EM coherence time v.s. turning radius.}
\label{fig_5}
\vspace{-0.25cm}
\end{figure}
In Fig. \ref{fig_5}, we plot the EM coherence time as a function of the turning radius $\rho$ and compare it with the results that neglect polarization mismatch or beam misalignment. The result obtained by ignoring beam misalignment is the EM coherence time calculated numerically from \eqref{N}. When $\rho$ is less than $10$ m, corresponding to the turning scenario, the trend of coherence time obtained by neglecting the beam misalignment effect closely matches the simulation result. As $\rho$ increases, beam misalignment becomes the primary factor influencing the channel temporal correlation. When $\rho$ exceeds $500$ m, the motion trajectory degenerates into linear displacement, and the EM coherence time obtained by neglecting polarization mismatch shows a deviation of less than 0.3 dB from the simulation result, indicating the diminished impact of polarization mismatch in the linear motion scenario.

\section{Conclusion}
In this paper, we have presented an analytical framework grounded in EM theory to derive the physically accurate expression for the EM coherence time. We have derived a closed-form expression for the EM coherence time for the turning scenario, which
provides insights into the impact of polarization mismatch on the mobile channel.
For the linear motion scenario, a closed-form expression for the lower bound on the EM coherence time has also been proposed as a practical parameter for protocol design.
Simulation results have revealed that the EM coherence time can be overestimated by up to one order of magnitude in the turning scenario by neglecting the impact of polarization mismatch.
This thus highlights the critical role of EM polarization effects in characterizing coherence time. Besides, we have demonstrated that the closed-form expressions for both scenarios match the simulation results, making them efficient alternatives to time-consuming numerical procedures.

\bibliographystyle{IEEEtran}
\bibliography{rece.bib, mybibliography.bib}

\end{document}